\documentclass[]{spie}  %>>> use for US letter paper
\usepackage{amsmath,amsfonts,amssymb}
\usepackage[colorlinks=true, allcolors=blue]{hyperref}
\usepackage{csquotes}

\usepackage{graphicx,caption,subcaption}
\usepackage{enumitem}
\graphicspath{{plots/}}

\title{Planet detection down to a few \textbf{\Large{$\mathbf{\lambda/D}$}: an RSDI/TLOCI approach to PSF subtraction}}

\author[a,b]{Benjamin L. Gerard}
\author[b,a]{Christian Marois}
\author[ ]{the GPIES team}
\affil[a]{University of Victoria, Department of Physics and Astronomy, 3800 Finnerty Rd., Victoria, V8P 5C2, Canada}
\affil[b]{National Research Council of Canada, 5071 West Saanich Rd, Victoria, V9E 2E7, Canada}
\authorinfo{Further author information: (Send correspondence to B. Gerard)\\B. Gerard.: E-mail: bgerard@uvic.ca}

\begin{document} 
\maketitle

\begin{abstract}
Most current high contrast imaging point spread function (PSF) subtraction algorithms use some form of a least-squares noise minimization to find exoplanets that are,  before post-processing, often hidden below the instrumental speckle noise. In the current standard PSF subtraction algorithms, a set of reference images is derived from the target image sequence to subtract each target image, using Angular and/or Simultaneous Spectral Differential Imaging (ADI, SSDI, respectively). However, to avoid excessive exoplanet self-subtraction, ADI and SSDI (in the absence of a strong spectral feature) severely limit the available number of reference images at small separations. This limits the performance of the least-squares algorithm, resulting in lower sensitivity to exoplanets at small angular separations. Possible solutions are to use additional reference images by acquiring longer sequences, use SSDI if the exoplanet is expected to show strong spectral features, or use images acquired on other targets. The latter option, known as Reference Star Differential Imaging (RSDI), which relies on the use of reference images that are highly correlated to the target image, has been ineffective in previous ground-based high contrast imaging surveys. The now $>$200 target reference library from the Gemini Planet Imager Exoplanet Survey (GPIES) allows for a detailed RSDI analysis to possibly improve contrast performance near the focal plane mask, at $\sim$2-7 $\lambda/D$ separations. We present the results of work to optimize PSF subtraction with the GPIES reference library using a least-squares algorithm designed to minimize speckle noise and maximize planet throughput, thus maximizing the planet signal to noise ratio (SNR). Using December 2014 51 Eri GPI data in the inner 100 mas to 300 mas annulus, we find no apparent improvement in SNR when using RSDI and/or our optimization scheme. This result, while still being investigated, seems to show that current algorithms on ADI+SSDI data sets are optimized, and that limited gains can be achieved by using a PSF archive.
\end{abstract}


\section{INTRODUCTION}
\label{sec: intro}

In the search to detect and characterize exoplanets by direct imaging, the best achievable contrast requires suppression effects from both a realtime adaptive optics (AO) and coronagraphic system as well post-processing of these images to remove the residual point spread function (PSF). Current state of the art PSF subtraction algorithms are limited in sensitivity at small angular separations from the on-axis PSF, close to the coronagraph's focal plane mask (FPM). Exoplanet population predictions suggest that with the current generation of high contrast imaging instruments, more detections are possible at smaller separations\cite{gpi_planets}, or alternatively there could be additional planets in existing data that could be seen with better post-processing sensitivity. It is suggested that the distribution of radial-velocity detected planets as a function of separation follows an inverse power law\cite{rv_planets}, continuing to wider separations accessible by direct imaging\cite{gdps}, and thus improved post-processing performance at smaller angular separations is of great interest to the exoplanet community. 

The main factor limiting PSF subtraction performance near the inner working angle (IWA) is the selection criteria for angular differential imaging (ADI)\cite{adi} and simultaneous spectral differential imaging (SSDI)\cite{ssdi1,ssdi2,ssdi3}. This selection criteria requires that reference images, used in some form of a least-squares-based\cite{loci, klip} PSF subtraction algorithm, have a certain planet signal threshold (known as ``aggressiveness'') to limit planet self-subtraction in the target image. The amount of field of view (FOV) rotation or spectral magnification required with ADI or SSDI, respectively, usually between $\sim$1.5 and 3 $\lambda/D$, limits fewer available references from the target sequence at smaller separations (i.e., with SSDI, for a flat spectrum, compared to a larger separation, diffraction at a smaller separation moves speckles a smaller absolute radial distance as a function of wavelength\footnote{For a spectrum with strong spectral features like methane absorption, the selection criteria will allow many more references to include in the least-squares at the peak vs. trough of the spectrum, thus ``counteracting'' the small angular separation selection bias.}, and with ADI, at a set FOV rotation, arc length decreases with decreasing radial separation). Thus, this ADI+SSDI selection effect limits the optimal PSF subtraction sensitivity to planet detection at small angular separations near the instrument IWA.

A solution to this problem is reference star differential imaging (RSDI), where PSF subtraction on a target image that may contain a hidden planet can access a large archive of ``planet-less'' references images. The key to increasing sensitivity at small IWA using RSDI, as with any PSF subtraction technique, is to use a set of reference images that are highly correlated to the target image. With this in mind, the archival legacy investigations of circumstellar environments (ALICE) pipeline\cite{alice} was recently developed for \textit{Hubble Space Telescope} (\textit{HST}) PSF subtraction, mostly to recover disks, but no ground-based first generation AO surveys, generally accessing a higher sensitivity and smaller IWA than \textit{HST}, have yet acquired enough data for RSDI in the $\sim$2-7$\lambda/D$ regime\footnote{Ground-based RSDI was initially attempted with pervious-generation generation high contrast imaging surveys beyond $\sim$7 $\lambda/D$, but with no performance gain, likely due to stability issues (e.g., Galicher et al. 2016, submitted)}.

With the next generation of high contrast imaging instrument surveys such as the Gemini Planet Imager Exoplanet Survey (GPIES)\cite{gpies}, we can test the performance of RSDI down to $\sim$2 $\lambda/D$. Our initial work on this topic acts as demonstrator for RSDI performance gain in this regime with future high contrast imaging survey instruments. In this paper, we present an analysis using GPIES to increase planet sensitivity near the IWA with a least-squares-based RSDI algorithm. In \S\ref{sec: lib} we describe our procedure used to create PSF library reference images, in \S\ref{sec: algo} we outline the specifics of our algorithm, in \S\ref{sec: results} we present the results of our algorithm applied to December 2014 GPI 51 Eri data\cite{51eri}, and in \S\ref{sec: conclusion} we summarize our work and consider possible future improvements.

This analysis is based entirely on GPI H band 51 Eri target sequence data from December 2014\cite{51eri} and an additional PSF library archive from the GPIES campaign through December 2015. We only consider performance close to the edge of the FPM ($\sim$125 mas)\cite{gpi_coron} in the inner 100 mas to 300 mas annulus. Matching the 51 Eri b detection, all of the following analysis is carried out using a methane (T8) dwarf spectrum. We use zero indexing to refer to frames and slices in the target sequence. We use a GPI pixel scale of 14.166 mas/spaxel, determined from all GPIES and lab astrometric data\cite{plate_scale}. We note that a similar RSDI procedure with GPIES was recently developed using the Karhunen-Lo{\`e}ve Image Projection (KLIP) algorithm\cite{klip}, mostly designed for broadband disk detection (M. Millar-Blanchaer et al., in prep) and broadband planet detection at wider separations (D. Vega et al, in prep), and so the work presented in this paper is complimentary, instead using a least-squares designed for small IWA planet detection and spectral extraction.

\section{PSF LIBRARY}
\label{sec: lib}

\begin{figure}[!h]
\begin{center}
\includegraphics[height=8cm]{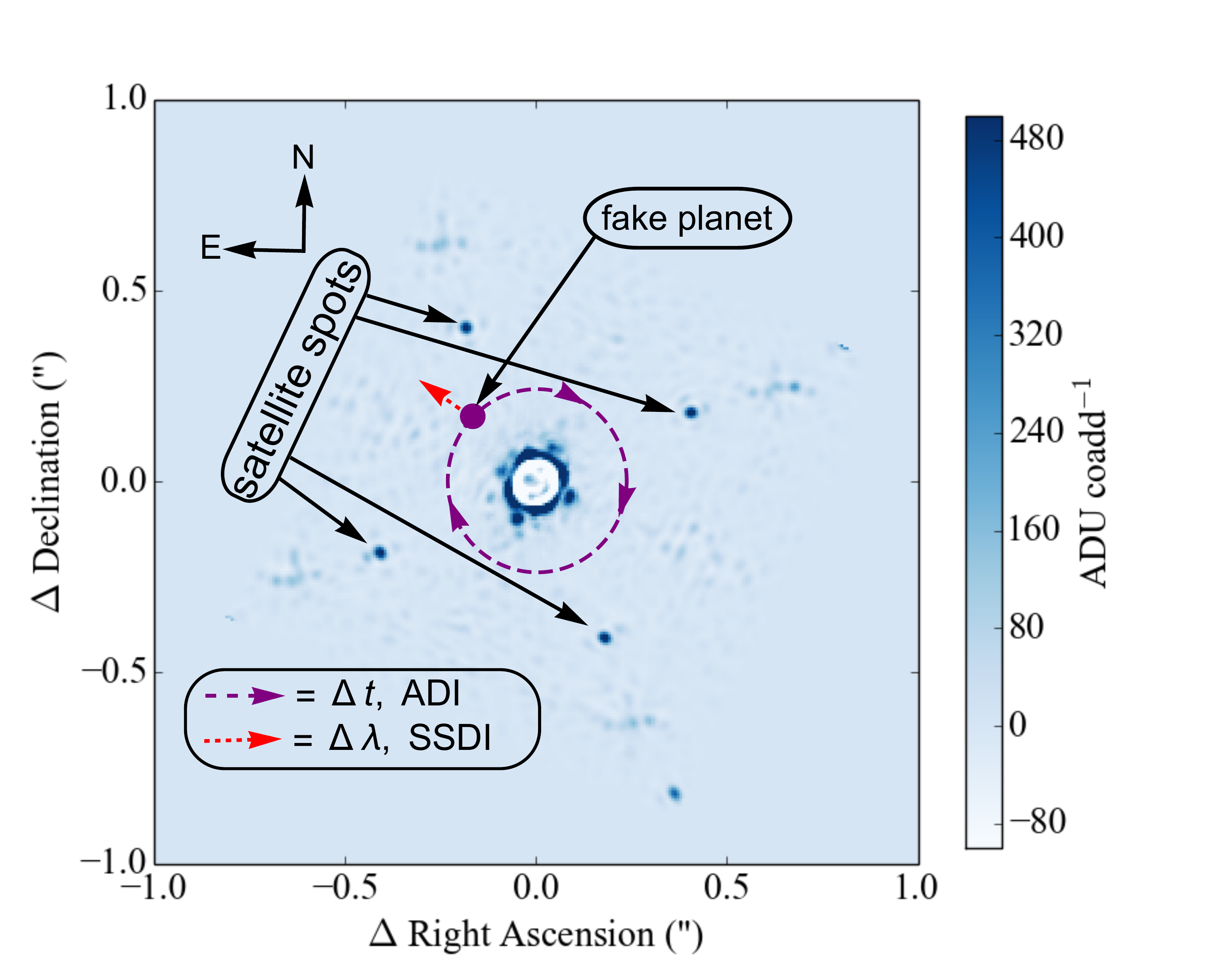}
\end{center}
\caption{A schematic diagram, from a GPIES target sequence at a single time and wavelength after steps \ref{step: mag} - \ref{step: filt} above, illustrating how creating the PSF archive images in step \ref{step: med} removes any planet signal by medianing across time and wavelength, where the planet position changes with respect to the telescope pupil due to ADI and SSDI.}
\label{fig: archive}
\end{figure} 

Figure \ref{fig: archive} shows a schematic of how we create the GPIES PSF library such that any potential planet is removed from the data. In each sequence of images for a given target, we use the datacubes produced from the GPI data reduction pipeline\cite{drp} so that all images at a given time and wavelength are
\begin{enumerate}
\item\label{step: mag} registered to a common center using cubic spline interpolation, magnified as a function of wavelength to align speckles (the first step in simultaneous spectral differential imaging, or SSDI), and flux normalized to flatten the stellar spectrum, all using the GPI satellite spot\cite{gpi_sat} positions,
\item\label{step: filt} high-pass filtered using a 11 by 11 pixel median boxcar filter to remove low spatial frequency noise, and
\item\label{step: med} median combined in both time and wavelength, removing any possible planet signal as illustrated in Figure \ref{fig: archive}.
\end{enumerate}
Any exoplanet will be removed after medianing all images in step \ref{step: med} because (1) the planet position in ADI observing mode changes azimuthally with time (i.e., without de-rotating the images to a common position angle), and (2) after step \ref{step: mag} the planet position changes radially with wavelength. Thus, for every target sequence we obtain one PSF library reference image, yielding a total of 207 available archive references.

\section{PSF SUBTRACTION ALGORITHM}
\label{sec: algo}

In this section we present our adaptation of the Speckle-Optimized Subtraction for Imaging Exolanets (SOSIE)\cite{sosie} and Template Locally Optimized Combination of Images (TLOCI)\cite{tloci} least-squares pipelines to use with optimized RSDI. In \S\ref{sec: tloci} we review the principles of SOSIE and TLOCI in application to our pipeline, in \S\ref{sec: available_refs} we explain our reference selection criteria based on image correlation, in \S\ref{sec: opt} we outline our algorithm to optimize planet SNR, and in \S\ref{sec: test_cases} we discuss additional optimization parameters. For all images in the 51 Eri target sequence, unless explicitly stated, we apply all the steps in the TLOCI algorithm\cite{tloci}, using an input T8 spectrum.

\subsection{TLOCI, SOSIE Architecture}
\label{sec: tloci}

\begin{figure}[!h]
\begin{center}
\includegraphics[width=10cm]{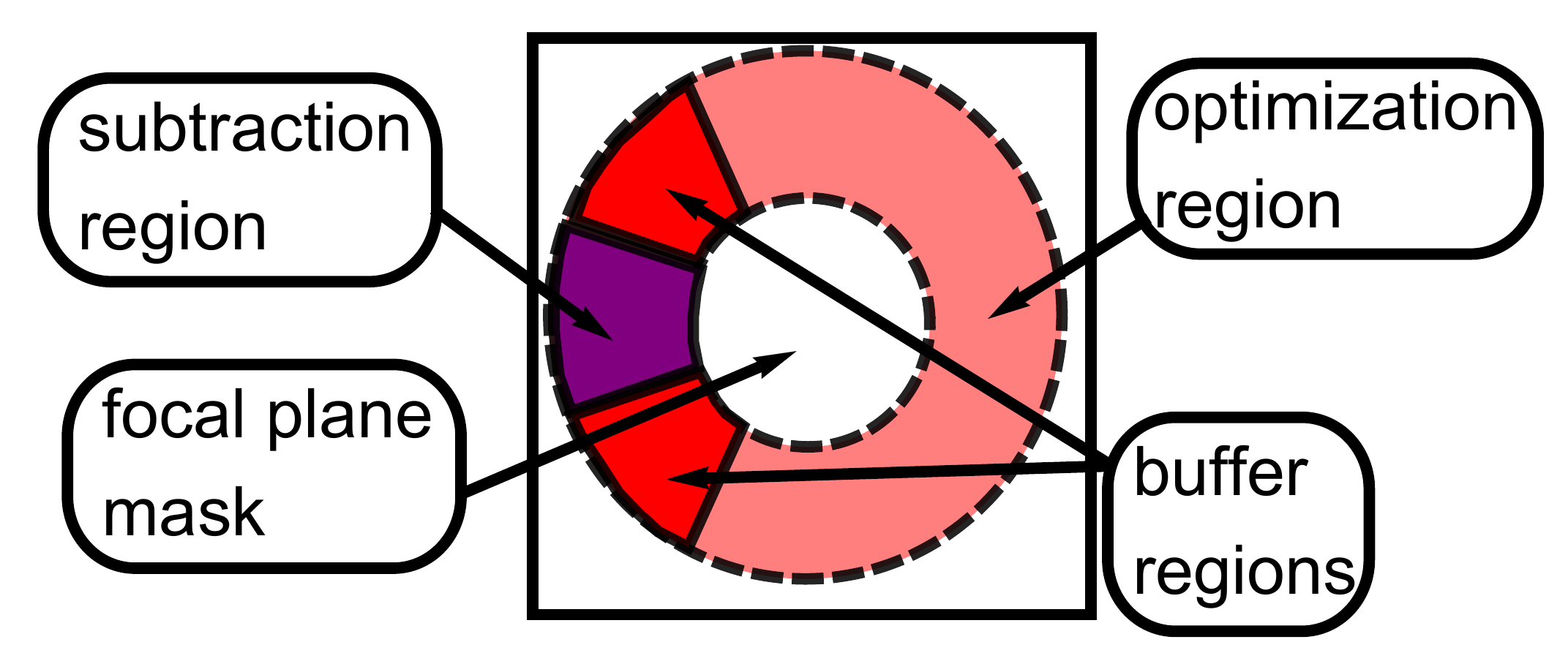}
\end{center}
\caption{A schematic of our chosen optimization and subtraction region geometry, adding buffer regions between the two. A least-squares algorithm\cite{loci} is run on the optimization region in the target image and set of references to generate subtraction coefficients that are then applied to the target image subtraction region.}
\label{fig: sosie}
\end{figure}

As in the original least-squares-based LOCI algorithm\cite{loci}, we define the region of interest to subtract the PSF as the ``subtraction region'' as well as a larger ``optimization region.'' The optimization region is used to obtain the least-squares subtraction coefficients from a region that does not contain any exoplanet light, thus fitting for the speckle noise while preventing the algorithm from fitting the planet signal. The subtraction coefficients are then applied to the subtraction region to minimize the noise but also preserve the planet signal. In order to obtain these subtraction coefficients, a least-squares algorithm is run using a correlated set of reference images (with the same geometry of optimization and subtraction regions as in the target image) in order to minimize the noise in the target image optimization region. In the LOCI algorithm, the subtraction region lies within the optimization region, causing the least-squares to fit any planet signal in the subtraction region, thus significantly affecting the algorithm throughput and planet SNR. In contrast, the SOSIE algorithm is designed to allow better planet throughput by
\begin{enumerate}
\item\label{step: mask} masking the subtraction region from the optimization region, and
\item\label{step: fm} correcting the remaining throughput reduction from self-subtraction by using a forward model (FM).
\end{enumerate}
Our choice of masking geometry in step \ref{step: mask} is illustrated in Figure \ref{fig: sosie}.
Any SOSIE implementation relies on the assumption that noise in the optimization region is spatially correlated with noise in the subtraction region, and thus works best with highly correlated data obtained in stable conditions. We chose the optimization region geometry in Figure \ref{fig: sosie}. Additionally, we chose to add a ``buffer region'' on either side of the optimization region, each extending azimuthally by the amount of target sequence FOV rotation (usually less than $\sim30^\circ$). The rationale for adding this buffer region is that if a planet lies at the edge of the subtraction region, the buffer region prevents the least-squares from fitting the planet light in an adjacent optimization region, potentially preventing throughput loss, although further testing of different SOSIE geometries, which are beyond the scope of this initial paper, are needed to better understand the sensitivity gain by adding a buffer region, as well as the validity of assuming the noise is azimuthally symmetric (see \S\ref{sec: conclusion}).

In step \ref{step: fm}, we use a TLOCI approach\cite{tloci} to generate spectrum-weighted template PSFs as a function of time and wavelength, applying a T8 template spectrum to select reference images at a given aggressiveness (see \S\ref{sec: available_refs}). We then apply the least-squares subtraction coefficients to the target and reference template PSFs, just as is done to subtract the real target images with a set of references, creating a residual noiseless FM to estimate the amount of self-subtraction and apply the appropriate throughput correction. Instead of using the FM flux value at the target PSF location to calculate a throughput correction, we use the peak FM flux value in an aperture centered on the target PSF location of width $\sim$$\lambda/D$ because the former can be biased by aggressive self-subtraction effects.
\subsection{Reference Image Selection}
\label{sec: available_refs}
\begin{figure}[!h]
\begin{center}
	\begin{subfigure}[b]{0.45\textwidth}
		\includegraphics[width=1.0\textwidth]{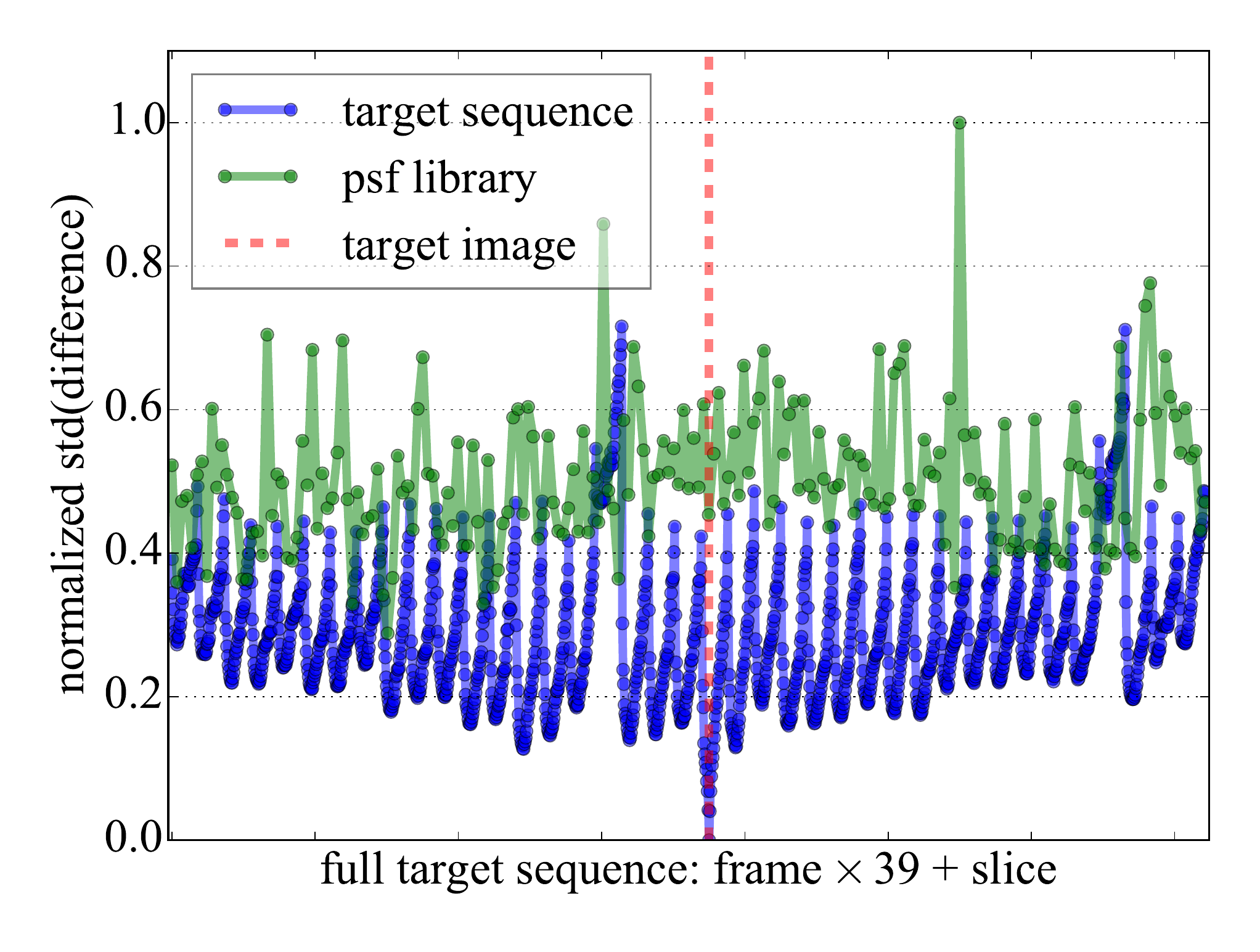}
		\caption{}
		\label{fig: a}
	\end{subfigure}
	\begin{subfigure}[b]{0.45\textwidth}
		\includegraphics[width=1.0\textwidth]{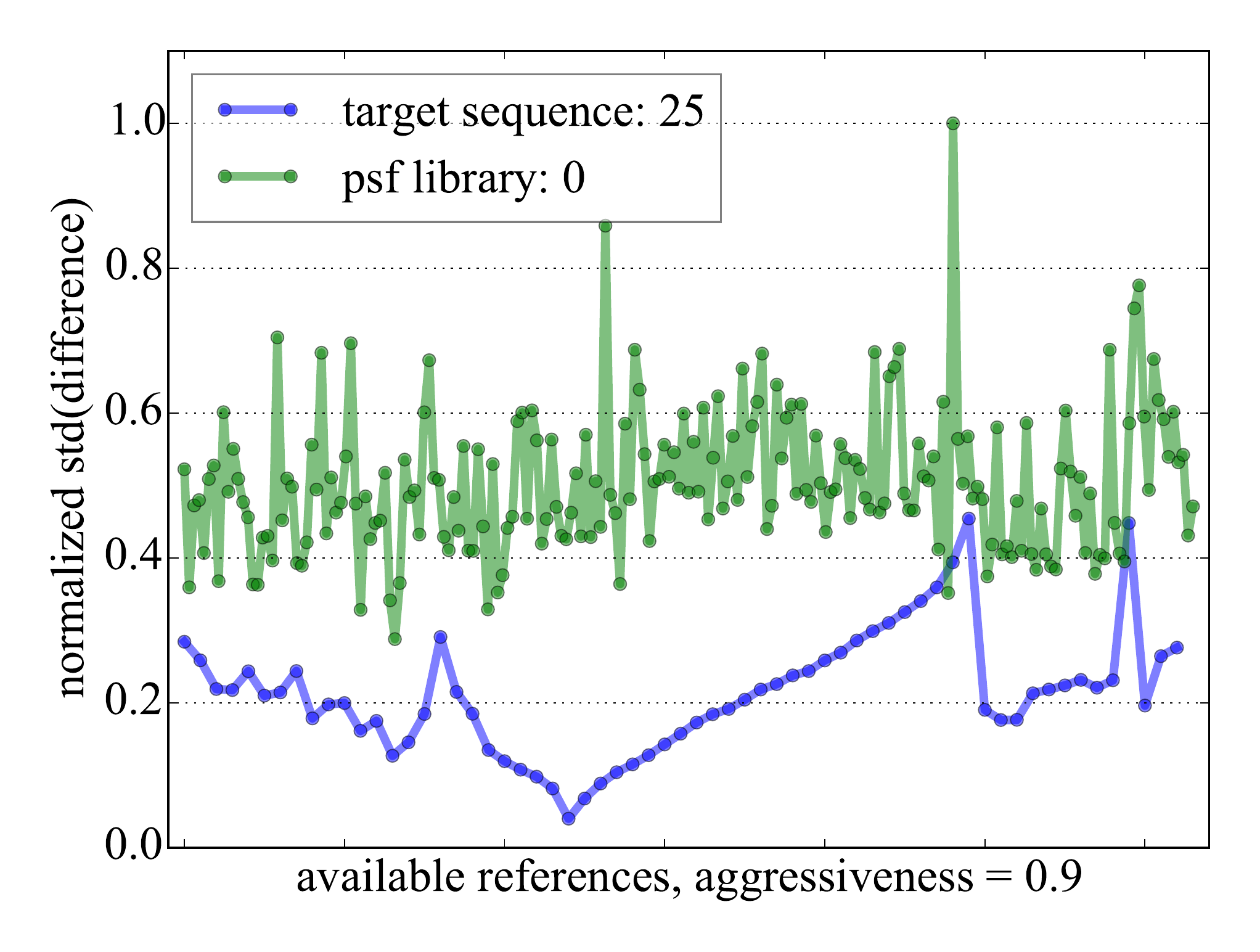}
		\caption{}
		\label{fig: b}
	\end{subfigure}
	\begin{subfigure}[b]{0.45\textwidth}
		\includegraphics[width=1.0\textwidth]{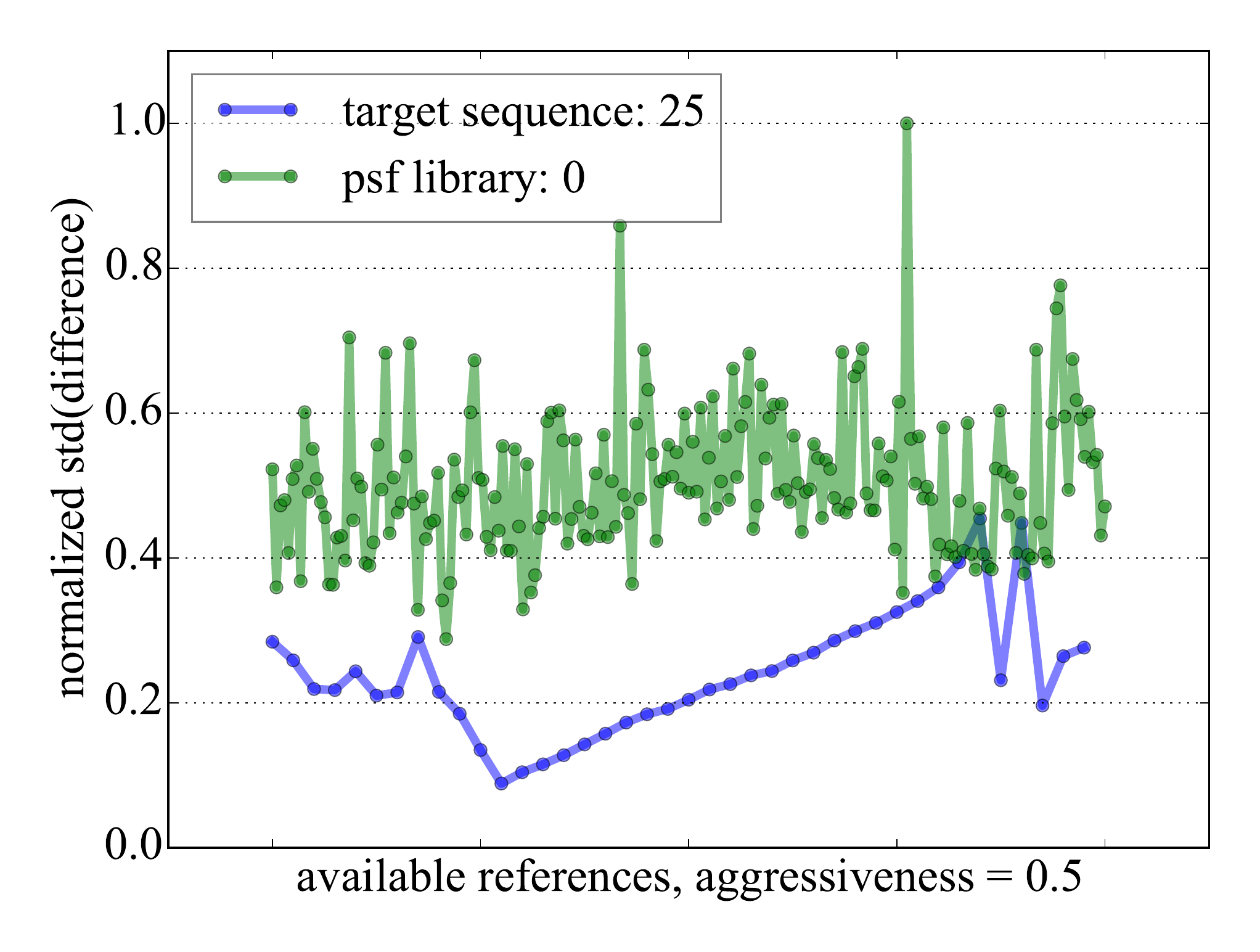}
		\caption{}
		\label{fig: c}
	\end{subfigure}
	\begin{subfigure}[b]{0.45\textwidth}
		\includegraphics[width=1.0\textwidth]{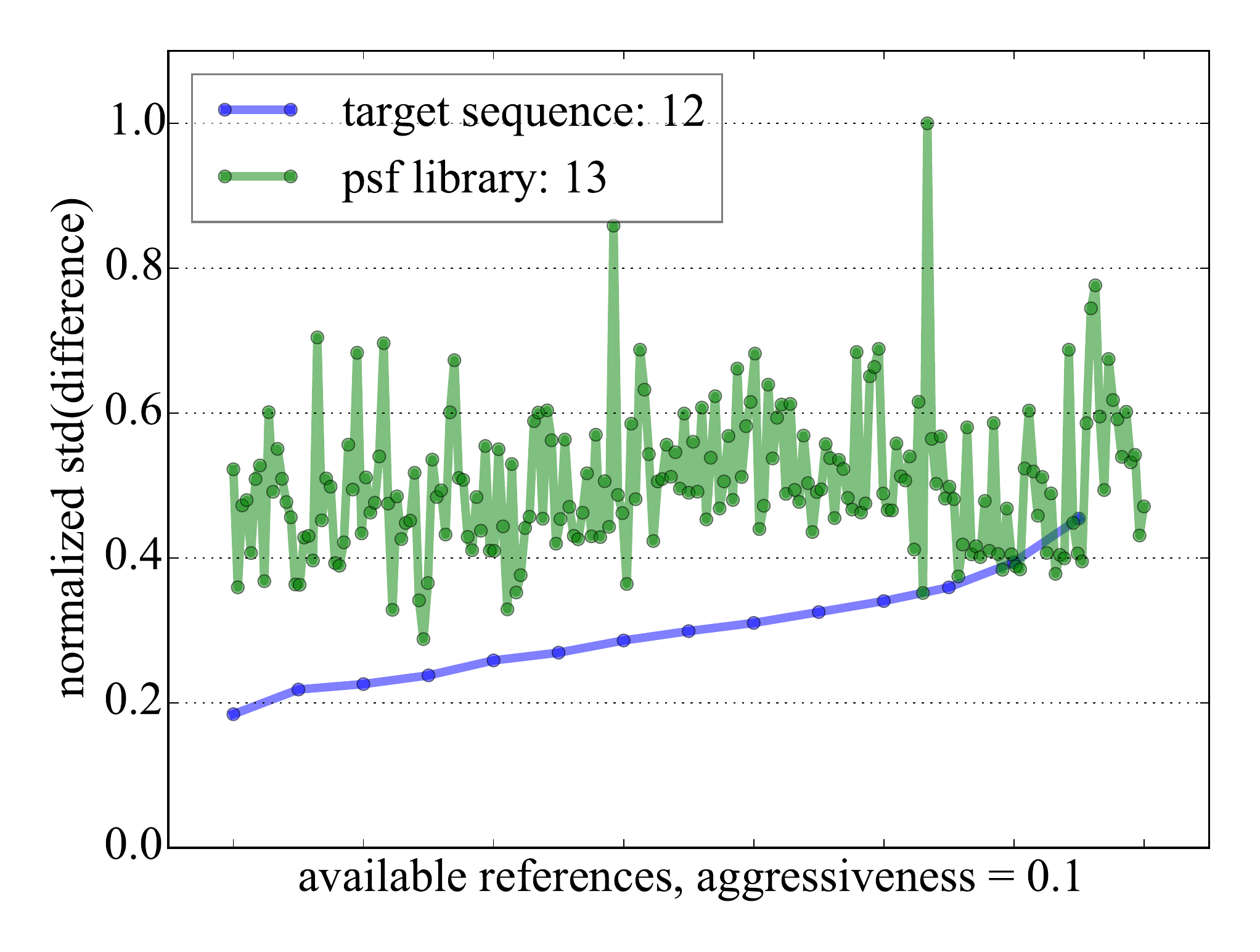}
		\caption{}
		\label{fig: d}
	\end{subfigure}
\end{center}
\caption{An example of how the correlation for a single target image (frame 20, slice 10) varies with aggressiveness. (a) Our correlation metric for an example target image no selection criteria, using the robust standard deviation of the difference between the target image and a given reference image, normalized to the highest value. As expected, the robust correlation converges to 0 at the sample target image. Using the reference selection criteria to chose the 25 most correlated images, an aggressiveness of 0.9 (b) and 0.5 (c) yields only one PSF library image to include in the set of references, whereas an aggressiveness of 0.1 (d) yields references mostly from the PSF library, suggesting that a gain in contrast by adding the PSF library is dependent on aggressiveness.}
\label{fig: corr}
\end{figure} 
Normalizing reference images to a given target image using the robust standard deviation\cite{robust_sigma}, we then used the robust standard deviation of the target-reference difference (Figure \ref{fig: a}) to quantify target image ``correlation,'' which converges to zero at the target image and is lower for more correlated reference images. Figures \ref{fig: b}, \ref{fig: c}, and \ref{fig: d} show the same robust correlation metric after applying the TLOCI selection criteria with aggressiveness 0.9, 0.5, and 0.1, respecively. An aggressiveness of 0.9 requires that for a given reference image, the cumulative FM flux in a $\lambda/D$ diameter aperture, centered on the target image FM location, be less than 90\% of the target image cumulative FM flux in that aperture (and for an aggressiveness of 0.5, less than 50\%, etc.). In addition to aggressiveness, we require that any reference for a given target image must be at either the same time or wavelength. This allows for a decreased overall computation time in order to determine the correlation of a given target image to all the available references. Although this limits the total available number images from which to select references, we know that images at the same time and wavelength are already $\sim$90\% correlated in wavelength and $\sim$50\% correlated in time\cite{tloci}. Thus, images at a different time and wavelength from the target image should accordingly decrease correlation, and so with this simplification we are still keeping the most correlated images.

Assuming use of the 25 most correlated reference images (after the TLOCI selection criteria to minimize self-subtraction effects; but see \S\ref{sec: opt}) to subtract the target image, Figures \ref{fig: b}, \ref{fig: c}, and \ref{fig: d} show that at high and medium aggressiveness, most references are from the target sequence (24 of 25), whereas at low aggressiveness, a greater number of the most correlated references come from the PSF library (11 of 25). For low aggressiveness, we can see that all the target sequence images are more correlated than any PSF library image, but this aggressiveness limits the total number of available references from the target sequence to 14, thus requiring use of the PSF library if we want to use 25 total references. This discrepancy as a function of aggressiveness, in both available number of correlated images from the target sequence and number of correlated images from the PSF library, suggests that adding the PSF library could improve contrast in at least some cases that require low aggressiveness, such as the spectral extraction of a companion.
\subsection{Optimization Algorithm}
\label{sec: opt}

Although the SOSIE and TLOCI algorithms have made improvements over the original LOCI algorithm to correct for throughput using masking and forward modelling, thereby effectively optimizing the planet SNR vs only minimizing the noise in the subtraction region, there are still (unaccounted for) free parameters which can change the planet SNR, including
\begin{enumerate}
\item\label{item: num_ref} number of references,
	\begin{itemize}[label={}]
	\item Too few references limits the diversity of the least-squares to optimally minimize the noise in the subtraction region, whereas too many reference images will overfit the noise in the optimization region, also causing a higher noise in the subtraction region.
	\end{itemize}
\item\label{item: aggr} aggressiveness,
	\begin{itemize}[label={}]
	\item A lower aggressiveness inherently requires selecting less correlated reference images from the target sequence but also causes less self-subtraction, whereas a higher aggressiveness will use more correlated images from the target sequence but cause more self-subtraction.
	\end{itemize}
\item\label{item: svd} singular value decomposition (SVD) cutoff,
	\begin{itemize}[label={}]
	\item Less correlated reference images can cause noise to propagate into the covariance matrix inversion step in the least-squares algorithm, effectively overfitting the noise in the optimization region\cite{sosie}, thus causing a higher noise in the subtraction region, similar to step \ref{item: num_ref}. This problem can be addressed by changing the SVD cutoff, which truncates the matrix inversion below a cutoff (singular) value.
	\end{itemize}
\item\label{item: geo} different optimization and subtraction region geometries.
\end{enumerate}
The core methodology of our algorithm is to address items \ref{item: num_ref} and \ref{item: aggr} by optimizing what we call the ``FM SNR.'' We also briefly address items \ref{item: svd} and \ref{item: geo} in \S\ref{sec: test_cases} and also item \ref{item: geo} in \S\ref{sec: conclusion}. 

To optimize the SNR as a function of number of reference images and aggressiveness in each subtraction region, we compute a grid search with our aforementioned SOSIE/TLOCI algorithm, varying the number of the most correlated references between 3 and 25 with a step size of 1 and varying the aggressiveness between 0.1 and 0.9 with a step size of 0.1. For each iteration, we compute the FM SNR by dividing the peak FM flux within a $\lambda/D$ diameter aperture centered on the target PSF location (as discussed in \S\ref{sec: tloci}) by the robust standard deviation in the subtraction region of PSF-subtracted image. Before calculating the noise in the subtraction region, we first convolve the original PSF-subtracted image with a $\lambda/D$ diameter Airy disk kernel\cite{astropy}, whose cumulative flux is normalized to one so that flux is conserved in the convolved image. This convolution step filters out pixel-to-pixel noise in order to distinguish $\lambda/D$-sized speckles as the noise floor. Finally, we adopt the parameters of number of references and aggressiveness for a given subtraction region that maximize the FM SNR. 

Throughout this process, in each least squares image, we found cases where negative least-squares subtraction coefficients generated an artificial increase in forward model flux, which bias the FM SNR optimiztion even though the planet is not actually brighter. The solution is to restrict the subtraction coefficients to be positive in the least-squares algorithm, also known as a non-negative least-squares (NNLS)\cite{nnls}, and so we use this algorithm instead of a regular least-squares for the rest of our analysis. We also chose to optimize the FM SNR as opposed to bootstrapping a signal into the real image because the former is a noiseless image that can isolate throughput effects from noise effects, unlike the latter. 

To test our algorithm performance in the rest of this paper, we will consider optimized PSF subtraction using only the target sequence (hereafter ``tar opt''), unoptimized PSF subtraction using only the target sequence (hereafter ``un-opt''), and optimized PSF subtraction using both the target sequence and PSF library (hereafter ``tar+lib opt''). We do not consider optimized PSF subtraction using only the PSF library, without the target sequence. A full reduction with only the PSF library would not do better than just the target sequence or target sequence + PSF library because Figure \ref{fig: corr} shows that the most correlated images are from the target sequence, independent of aggressiveness, and that these also outnumber the most correlated images to include from from the PSF library, so that on average for a given target image, the given aggressiveness is such that the majority of most references will be from the target sequence.

\begin{figure}[!h]
\begin{center}
	\begin{subfigure}[b]{0.45\textwidth}
		\includegraphics[width=1.0\textwidth]{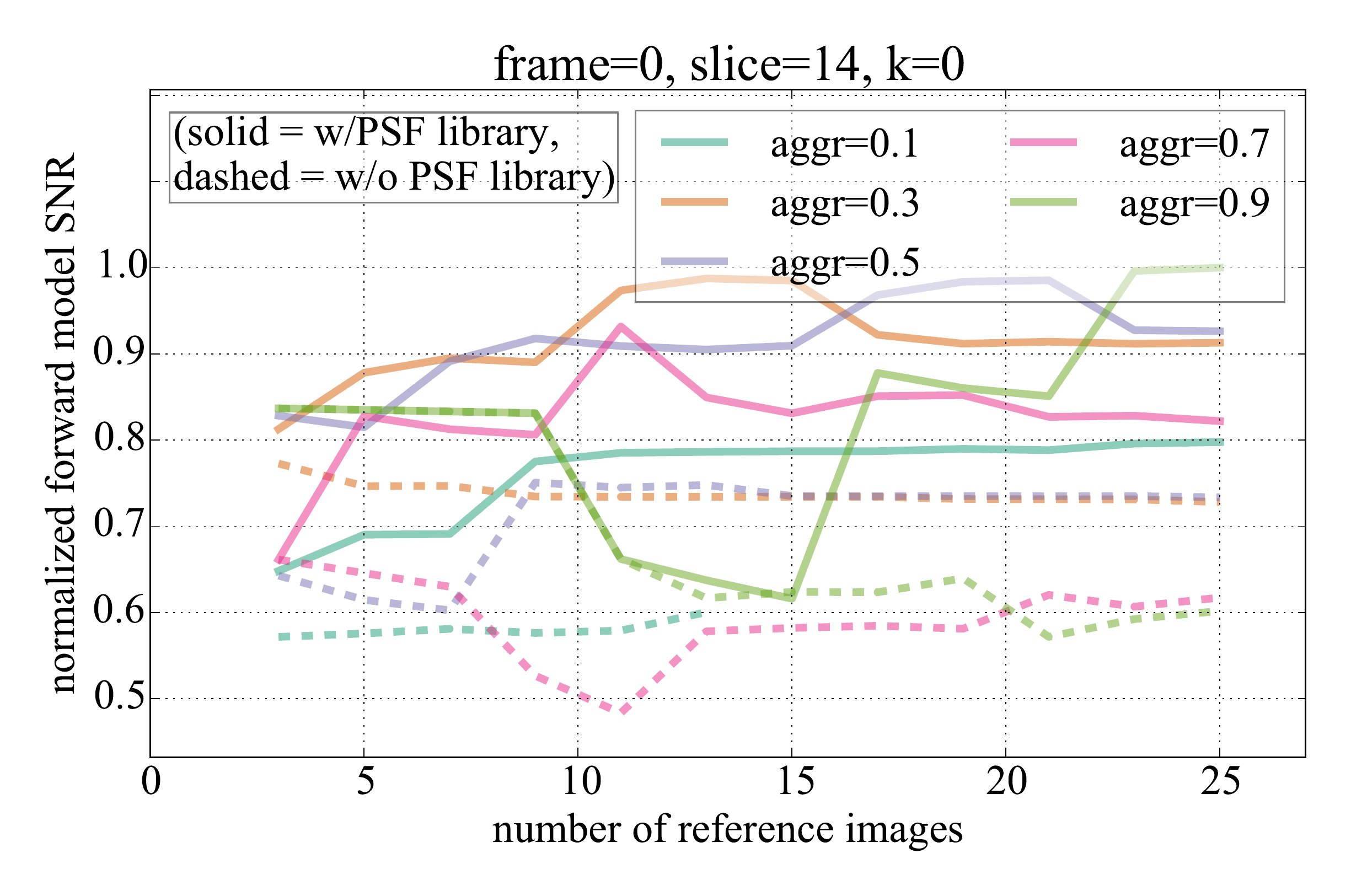}
		\caption{}
		\label{fig: aa}
	\end{subfigure}
	\begin{subfigure}[b]{0.45\textwidth}
		\includegraphics[width=1.0\textwidth]{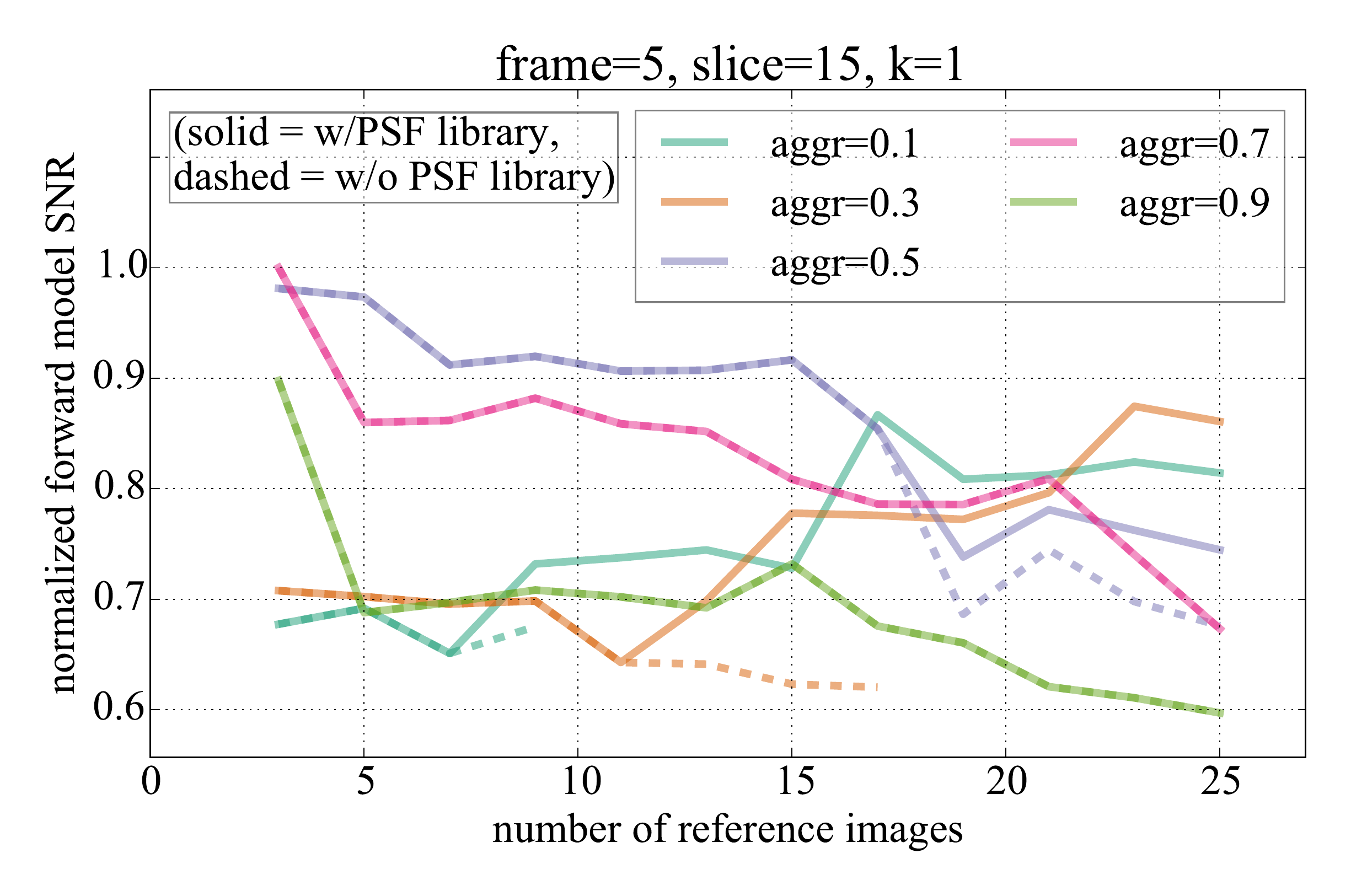}
		\caption{}
		\label{fig: bb}
	\end{subfigure}
	\begin{subfigure}[b]{0.45\textwidth}
		\includegraphics[width=1.0\textwidth]{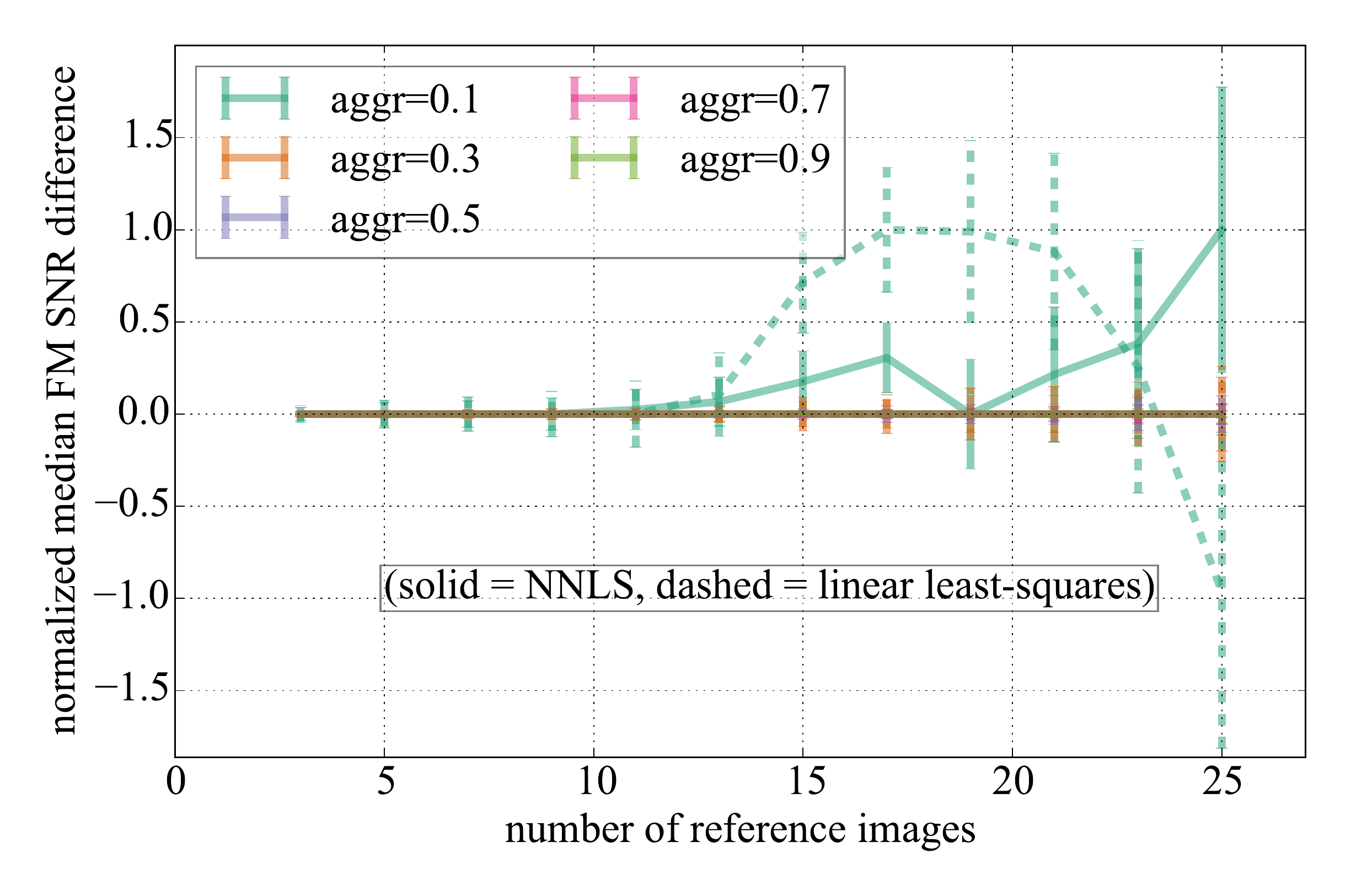}
		\caption{}
		\label{fig: cc}
	\end{subfigure}
	\begin{subfigure}[b]{0.45\textwidth}
		\includegraphics[width=1.0\textwidth]{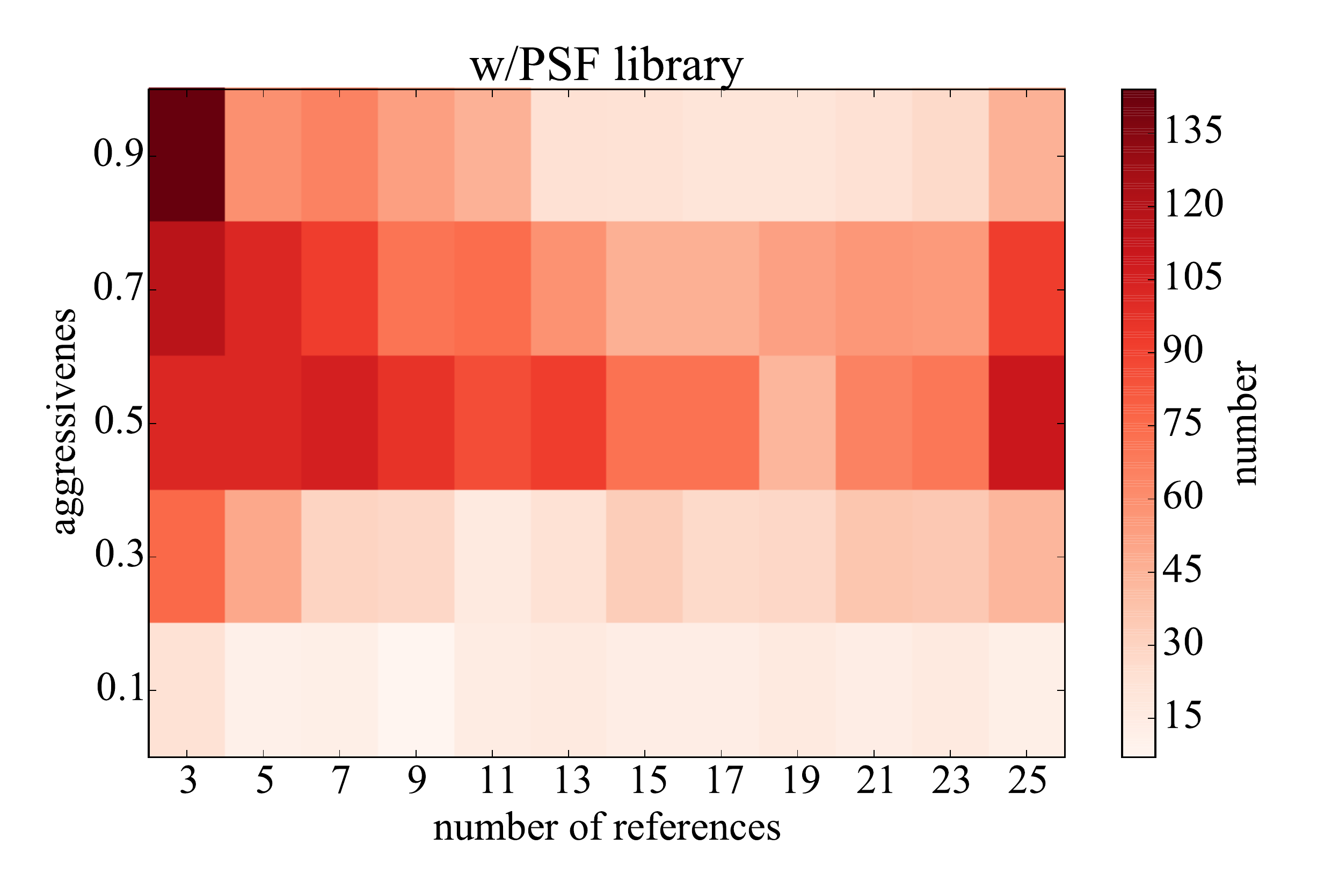}
		\caption{}
		\label{fig: dd}
	\end{subfigure}
\end{center}
\caption{(a)-(b) FM SNR as a function of number of reference images and aggressiveness, with and without the PSF library (``tar+lib opt'' and ``tar opt,'' respectively), for two test cases. The variable ``k'' refers to the specific subtraction region considered, for which, using two subtraction regions across the full annulus, k=0 represents the lower half of the annulus and k=1 represents the upper half. (c) the difference in FM SNR values between the tar+lib opt and tar opt codes for each individual least-squares image, averaged across the full sequence, as a function of number of reference images and aggressiveness. Error bars are determined from the standard deviation of each averaged FM SNR difference divided by the square root of the number of available values at that given number of references and aggressiveness. The non-negative least-squares (NNLS) algorithm is shown in solid for comparison with the standard linear lease-squares algorithm, shown as dashed lines. A positive value on the y-axis indicates that adding the PSF library improves the FM SNR. (d) A two-dimensional histogram of the optimal values for aggressiveness and number of references as determined from FM SNR optimization across the full target sequence using the tar+lib opt code. A higher number of counts in a given bin means that more target images are optimized at the corresponding number of references and aggressiveness compared to other bins. }
\label{fig: opt}
\end{figure}

Two examples comparing our various optimization schemes are shown in Figure \ref{fig: opt}, along with the mean behavior across the full target sequence.
Figure \ref{fig: opt} illustrates a number of important concepts that we discuss below:
\begin{itemize}
\item In Figure \ref{fig: aa}, the FM SNR is consistently higher in the tar+lib opt code than in the tar opt code (i.e., for a given color, the solid lines generally lie above the dotted lines) by up to a factor of $\sim$2, suggesting that adding the PSF library should show significant improvement in SNR, relatively independent of aggressiveness and number of references. However, this was a specially picked case where we found an unusually high number of correlated PSF library images to the target image.
\item Figure \ref{fig: bb} is a much more typical example FM SNR optimization. In this case, we can see that:
	\begin{enumerate}
	\item There is a $\sim$10-20 \% improvement in FM SNR when using the PSF library, but only at medium to low aggressiveness.
	\item without using the PSF library, a low aggressiveness selection criteria may provide less than 25 available references from the target sequence, meaning that the number of references can only be optimized out to less than this value. This effect agrees with Figure \ref{fig: corr}, showing that at a lower aggressiveness there are less available references from the target sequence.
	\end{enumerate}
The two points above suggest that adding the PSF library is most effective at low aggressiveness when there are  less available references from the target sequence. High aggressiveness cases do not show any improvement because there are always enough available references from the target sequence that are all more correlated with the target image than any PSF library image.
\item Figure \ref{fig: cc} shows that
	\begin{enumerate}
	\item adding the PSF library does on average improve FM SNR throughout the full sequence, but only at low aggressiveness and more than $\sim$10 references, consistent with Figure \ref{fig: bb}.
	\item at low aggressiveness, the negative coefficients in a linear least-squares algorithm are biasing the tar+lib opt FM to performance that is both better and worse than tar opt code, illustrating that using  a NNLS algorithm should yield a more consistent performance improvement in this regime.
	\end{enumerate}
\item Figure \ref{fig: dd} shows the diversity of optimal parameters for aggressiveness and number of references across the full sequence. This suggests that the standard approach to PSF subtraction, which uses a set number of references and aggressiveness across the full sequence (i.e. the un-opt code), should not perform as well as our optimized approach.
\end{itemize}
\subsection{Additional Parameters}
\label{sec: test_cases}
In order to choose an optimal least-squares SVD cutoff for matrix inversion (SVD$_\text{cutoff}$)\footnote{The SVD cutoff was only investigated for matrix inversion in our testing with a linear least-squares algorithm (but see \S\ref{sec: conclusion}). We did not consider optimizing the matrix inversion in the NNLS algorithm.} and number of subtraction regions around the full annulus ($n$), we ran a grid search with $10^{-1}<\text{SVD}_\text{cutoff}<10^{-7}$ (with step size of $\Delta$log$_{10}\left(\text{SVD}_\text{cutoff}\right)=-1$) and $2<n<5$ (using a step size of $\Delta{n}=1$) for a number of test cases at a fixed number of references and aggressiveness, using the same subtraction region geometry as in Figure \ref{fig: opt}. In a number of test cases for single target images, we generally found a maximum FM SNR at $\text{SVD}_\text{cutoff}=10^{-3}$ and $n=2$, and so we adopt these parameters for the rest of this paper (but see \S\ref{sec: conclusion}).

Relatedly, in order to run this optimization routine on the full target sequence in a reasonable amount of time ($\sim$24 hours on our 2.7 GHz processor, running in serial), we simplified the above optimization routine by binning the step size in number of reference images to two instead of one and in binning in aggressiveness to 0.2 instead of 0.1, finding no evidence of a significant decrease in forward model SNR from these effects for multiple test cases.
\section{RESULTS}
\label{sec: results}
\begin{figure}[!h]
\begin{center}
\includegraphics[height=11cm]{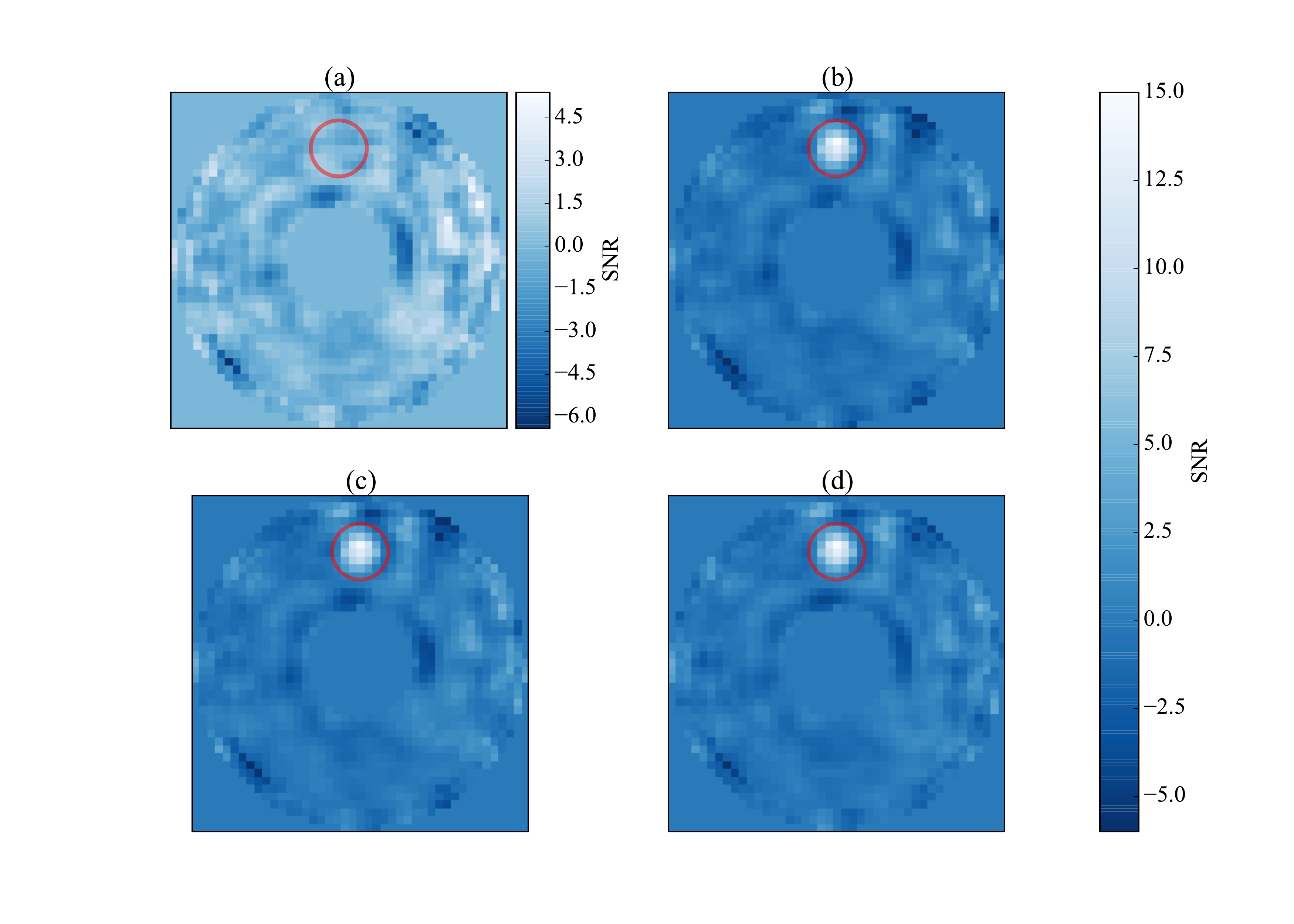}
\end{center}
\caption{SNR maps for (a) the optimized code using the PSF library, (b) bootstrapping with the optimized code using the PSF library, (c) bootstrapping with the optimized code using only the target sequence, and (d) bootstrapping with the un-optimized code (20 reference images, aggressiveness = 0.5 throughout the full sequence) using only the target sequence. The red circle shows the bootstrap location. The bootstrap SNR in (b), (c), and (d) is 12.55, 12.45, and 12.95, respectively.}
\label{fig: results}
\end{figure} 
\begin{figure}[!h]
\begin{center}
\includegraphics[width=1\linewidth]{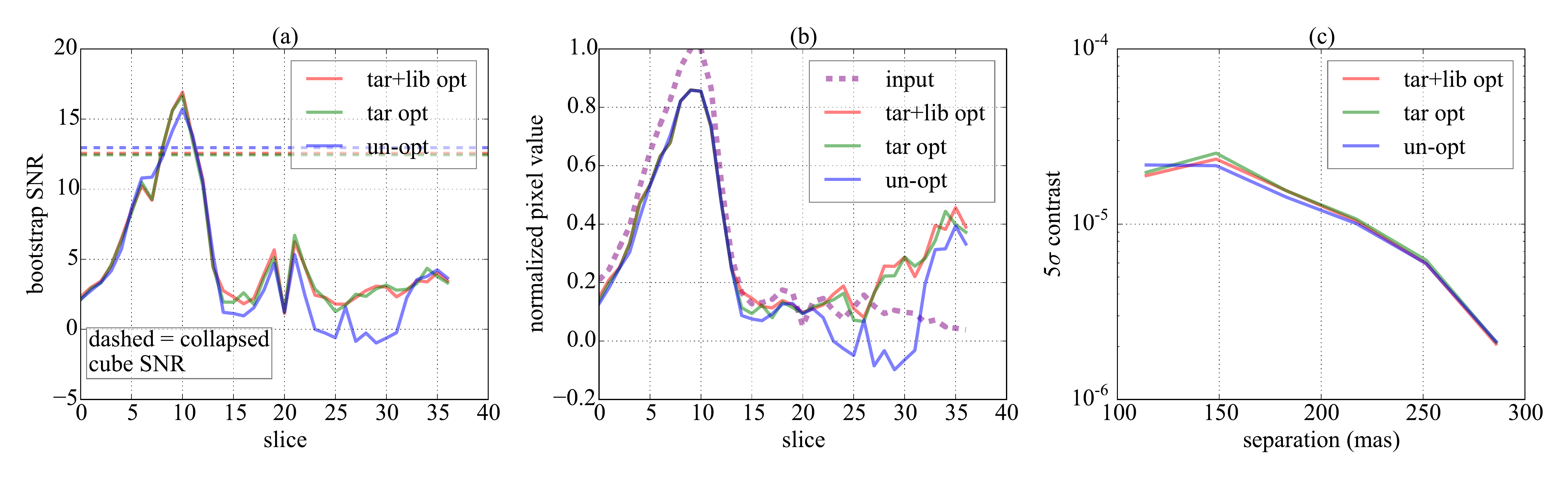}
\end{center}
\caption{Performance after bootstrapping a simulated methane planet using our three different codes: optimized using the target sequence and PSF library (tar+lib opt), optimized using only the target sequence (tar opt), and un-optimized (20 reference images, aggressiveness = 0.5 throughout the full sequence) using only the target sequence (un-opt). (a) Bootstrap SNR as a function of wavelength, where the dashed lines of the same color show the corresponding collapsed cube bootstrap SNR as in Figure \ref{fig: results} (b)-(d), indicating that neither our FM SNR optimization scheme nor using the PSF library improves performance. (b) Output bootstrap signal normalized to the peak value of the input bootstrap signal, indicating that all algorithms more or less recover the input signal, although the methane absorption band is bootstrapped well below the PSF-subtracted noise floor and thus biased by any brighter speckles, and the $\sim$20\% discrepancy at the T8 peak suggests that further work is needed to understand why our FM throughput correction is not performing as well in this regime. (c) 5$\sigma$ contrast as a function of separation on the collapsed, back-rotated, bootstrapped images, again showing that all codes have a similar performance.}
\label{fig: boot_results}
\end{figure} 

Results after running our tar+lib opt, tar opt, and un-opt codes on the full 51 Eri sequence, each of which include a run with and without a bootstrapped methane planet, are shown in Figure \ref{fig: results}. Each image is rotated to north up (ADI) and demagnified (SSDI) before median combining in time to create a final PSF-subtracted datacube. The datacube is then collapsed across wavelength with a weighted mean using the input spectrum\cite{tloci}. When running the un-opt code, we chose to use the ``typical'' parameters applied throughout the full sequence (independent of the results in Figure \ref{fig: dd}): 20 reference images and aggressiveness = 0.5. For the bootstrap reductions, we add the satellite spot PSF into the raw images, normalized to the robust standard deviation in slice 9 (the peak of a T8 spectrum) of the non-bootstrap tar+lib opt collapsed cube within a 2 pixel-wide annulus around the bootstrap radial separation (200 mas), multiplied by 8, which represents an ``$\sim$8$\sigma$'' detection. For a methane planet, the bootstrap signal is also a function of wavelength, and so we additionally normalize the bootstrap signal to the mean of the normalized input spectrum (accordingly, in the final time-collapsed cube the T8 peak should be above 8$\sigma$ and the methane absorption band should be below 8$\sigma$, and thus the bootstrap signal in the wavelength-collapsed image should be detected at $\sim$8$\sigma$). 

To compute the SNR maps in Figure \ref{fig: results}, we create a ``back-rotated'' time-collapsed datacube, where each frame is rotated by its amount of FOV rotation, but in the opposite direction of north up, thus medianing out any planet signal that is no longer spatially aligned, but preserving the radial noise characteristics. Within a given 2 pixel-wide annulus, the signal values in the correctly rotated north-up, time-collapsed image are then divided by the robust standard deviation in the same annulus of the corresponding back-rotated image, repeated over all annuli of the 100 to 300 mas region to produce the SNR maps in Figure \ref{fig: results}. 

None of the non-bootstrapped codes show an obvious point source detection above 5$\sigma$ in the inner annulus. We find a bootstrap SNR for the tar+lib opt, tar opt, and un-opt codes of 12.55, 12.45, and 12.95, respectively. These results suggest that neither our FM optimization scheme nor adding the PSF library actually improves planet SNR. The latter is consistent with the average results in Figures \ref{fig: cc}, which shows that on average the PSF library shows no improvement except in cases of low aggressiveness, for which Figure \ref{fig: dd} shows that most of the sequence is optimized at higher aggressiveness. However, the lack of improvement from our FM SNR optimization scheme is puzzling, and discussed further below as well as in \S\ref{sec: conclusion}.

Figure \ref{fig: boot_results} shows the bootstrap spectral performance and contrast curves for our three different PSF subtraction codes. The left panel shows the output bootstrap SNR spectrum (wavelength increases with increasing slices, from 0 to 36 across H band\cite{gpi_ifs}), confirming the collapsed cube results in Figure \ref{fig: results} as a function of wavelength: neither our FM SNR optimization scheme nor adding the PSF library shows any improvement in bootstrap SNR over existing methods. The middle panel shows the input and output bootstrap signals as a function of wavelength, normalized to the peak input signal, indicating that in all codes our noiseless FM throughput correction mostly recover the input signal. In the methane absorption band there is some discrepancy between the input and recovered signal, but in this low SNR regime the bootstrap signal is well below the noise and may be biased by brighter speckles. However, around the T8 peak, all codes are $\sim$20\% dimmer than the input signal. This throughput loss suggests that our FM throughput correction is not performing well in this regime, and further work is needed to understand why (\S\ref{sec: conclusion}). The right panel shows the 5$\sigma$ contrast as a function of position. In order to remove bias from the bootstrap signal, we calculate contrast on the bootstrapped, collapsed, back-rotated images. 
\section{CONCLUSION}
\label{sec: conclusion}
ADI and SSDI limit PSF subtraction sensitivity to detect and characterize planets at small angular separations. This problem can be addressed using RSDI. RSDI on ground-based high contrast imaging instruments at $\sim$2-7 $\lambda/D$ has only recently become possible with sufficient campaign data from the Gemini Exoplanet Imaging Survey (GPIES)\cite{gpies}. Our main conclusions are as follows:
\begin{itemize}
\item We have developed a new method of optimized RSDI PSF subtraction using the SOSIE\cite{sosie} and TLOCI\cite{tloci} least-squares formalism, which includes
	\begin{enumerate}
		\item reference image selection based on an input spectral template and robust correlation to the target image,
		\item an ensemble of PSF library reference images that removes any planet signal through medianing, many of which are correlated enough to images in the target sequence to allow using RSDI,
		\item an improved planet throughput compared to the LOCI\cite{loci} formalism, based on an optimization and subtraction region masking scheme and a subsequent forward model (FM) throughput correction, and
		\item an optimization routine designed to maximize the planet SNR as a function of the number of reference images and aggressiveness.
	\end{enumerate}
\item When running this PSF subtraction routine on the inner 100 to 300 mas annulus of the GPI December 2014 51 Eri dataset\cite{51eri}, we find
	\begin{enumerate}
		\item no obvious planet detection,
		\item when bootstrapping a fake methane planet into the raw datacubes, there is no apparent gain in planet SNR when adding the PSF library and/or using our FM optimization scheme compared to current non-RSDI-based PSF subtraction methods.
	\end{enumerate}
\end{itemize}
These results present the first attempt to improve planet SNR using RDSI in the $\sim$2-7 $\lambda/D$ regime. This method should be further explored in the context of current and future high contrast imaging survey instruments working near the diffraction limit. Future work on this initial study will proceed for a number of different topics. However, for any further adjustments to the optimized code, the first step is to parallelize the now serial Python-based optimization code so that it can reduce the full sequence in only a few hours on our 16 core machine rather than $\sim$24 hours. Afterwards, there are a number of different possible avenues to explore:
\begin{itemize}
\item Understand the discrepancy between input and output bootstrap signal near the T8 peak, perhaps originating from problems with applying our FM throughput correction in this regime. This throughput loss may also be affecting our FM SNR optimization scheme.
\item Test performance of the three codes using additional spectral templates, such as a DUSTY\cite{dusty} spectrum, which should be less effective than a methane spectrum when using SSDI due to the lack of spectral features, thus more sensitive to increased performance with RSDI.
\item Test performance of different optimization and subtraction region geometries, e.g., using all or portions of the adjacent, more outer annulus and/or portions of the inner annulus to define the optimization region geometry, still not overlapping with the subtraction region. The rational here is that there could be a more optimal geometry that samples the noise of the subtraction region rather than assuming azimuthal symmetry in the same annulus (e.g., assuming radial symmetry).
\item To further investigate the performance tradeoff between a linear least-squares and a NNLS, include a loop in the linear least-squares to optimize the SVD cutoff. Although we found that using two subtraction regions consistently gave the highest FM SNR, we found that the optimal SVD cutoff varied between $10^{-3}$ and $10^{-4}$. We did not include a loop to optimize this parameter due to computational limits, which would have at minimum doubled the overall computation time in serial. However, there may be little or no gain from adding such a loop, since optimizing the SVD cutoff is similar to optimizing the number of references.
\item Create PSF library images by medianing only across wavelength. This could allow for a greater ensemble of references from which to choose. However, medianing only in wavelength is also less effective at medianing out the planet flux in the inner annulus, and so there may be additional throughput effects from this. The rational here is that by median collapsing in time and wavelength, we could be missing a potentially more correlated PSF library image due to time instability. 
\item Run the tar+lib opt code on the all of the GPIES campaign data acquired thus far to search for any undetected planets already in the existing data, using multiple spectral templates. Most datasets are far less ideal than our 51 Eri data set, particularly in stability and FOV rotation, for which the former and the latter can be improved by using an SNR optimization scheme and a PSF library, respectively. Thus, RSDI and/or FM SNR optimization may show more improvement on other targets.
\end{itemize}

\acknowledgments % equivalent to \section*{ACKNOWLEDGMENTS}       
 
We gratefully acknowledge research support of the Natural Sciences and Engineering Council (NSERC) of Canada. The GPI project has been supported by Gemini Observatory, which is operated by AURA, Inc., under a cooperative agreement with the NSF on behalf of the Gemini partnership: the NSF (USA), the National Research Council (Canada), CONICYT (Chile), the Australian Research Council (Australia), MCTI (Brazil) and MINCYT (Argentina).
%
% References
\bibliography{refs} % bibliography data in report.bib
\bibliographystyle{spiebib} % makes bibtex use spiebib.bst

\end{document}